\documentclass[submission,copyright,creativecommons]{eptcs}
\providecommand{\event}{INFINITY 2011} 
\usepackage{breakurl}             

  \newcommand{\com}[1]{$\Leftarrow$ {Card1} $\Rightarrow$}

\newtheorem{definition}{Definition}
\newtheorem{theorem}{Theorem}
\newtheorem{example}{Example}
\newtheorem{lemma}{Lemma}
\newtheorem{proposition}{Proposition}

\usepackage{xcolor}
\usepackage{listings}
\usepackage{amsmath}
\usepackage{amssymb}
\usepackage{calc}
\usepackage{rotating}
\usepackage{subfigure}
\usepackage[ruled,vlined,linesnumbered]{algorithm2e}

\usepackage{pgf}

\newcommand{\todo}[1]{{\bf \textcolor{red}{[Card1]}}}

\lstdefinelanguage{pha}
{
morekeywords=[1]{automaton, contr_var, synclabs, loc, while, wait, when, sync, \
do, goto, end, initially},
alsoletter={=},
morekeywords=[2]{==,=},
keywordstyle=[2]{\tt},
sensitive,
morecomment=[l]{//},
morecomment=[s]{/*}{*/},
morestring=[b]'',
escapeinside={/*@}{@*/},
basicstyle=\sffamily\footnotesize,
breaklines=true,
xleftmargin=0.8cm,
numbers=left,
mathescape=true
}[keywords,comments,strings]

\lstdefinelanguage{imi}
{
morekeywords=[1]{automaton, var, clock, discrete, parameter, analog, synclabs, \
loc, while, wait, when, sync, do, goto, end, initially},
alsoletter={=},
morekeywords=[2]{==,=},
keywordstyle=[2]{\tt},
sensitive,
morecomment=[l]{//},
morecomment=[s]{/*}{*/},
morestring=[b]'',
scapeinside={(@}{@)},
basicstyle=\sffamily\footnotesize,
breaklines=true,
xleftmargin=0.8cm,
numbers=left,
mathescape=true
}[keywords,comments,strings]

\def\titlerunning{Synthesis of Switching Rules
for Ensuring Reachability Properties of  Sampled Linear Systems}
\def\authorrunning{L. Fribourg, B. Revol \& R. Soulat}
\begin{document}

\title{Synthesis of Switching Rules
for Ensuring Reachability Properties of  Sampled Linear Systems}

\author{Laurent Fribourg \institute{LSV, ENS Cachan $\&$ CNRS, 94230 Cachan, France \email{fribourg@lsv.ens-cachan.fr}}
 \and Bertrand Revol \institute{SATIE, ENS Cachan $\&$ CNRS, 94230 Cachan, France \email{bertrand.revol@satie.ens-cachan.fr} }
\and Romain Soulat \institute{LSV, ENS Cachan $\&$ CNRS, 94230 Cachan, France \email{soulat@lsv.ens-cachan.fr} }
}

\maketitle

\begin{abstract}
We consider here  systems with piecewise linear dynamics
that are periodically sampled
with a given period $\tau$. 
At each sampling time, the mode of the system, i.e.,
the parameters of the linear dynamics,
can be switched, according to a switching rule. Such systems can be modelled as a special form of hybrid automata, called ``switched systems'', that are automata with an \textit{infinite} real state space.
The problem is to find a switching rule
that guarantees the system to still be
in a given area $V$ at the next sampling time, and so on indefinitely.
In this paper, we will consider two approaches:
the {\em indirect} one that abstracts the system under the form of a finite discrete event system, and the {\em direct} one that works on the
continuous state space.\\
Our methods rely on previous works, but we specialize them
to a simplified context 
(linearity, periodic switching instants, absence of control input),
which is motivated by the features of a focused case study:
a DC-DC boost converter built by electronics laboratory SATIE (ENS Cachan).
Our enhanced methods allow us to treat successfully this real-life example.

\end{abstract}
\section{Introduction}

We are interested here in finding rules for switching the modes
of (piecewise) linear systems in order to make the variables
of the system stay within the limits of given area $V$.
The systems that we consider are periodically sampled
with a given period $\tau$. Between two sampling times,
the variables follow a certain system of linear  differential equations,
corresponding to a {\em mode} among several other ones. At each sampling time, the mode of the system
can be switched. Such systems can be modelled as a special form of hybrid automata, called ``switched systems'', that are automata with an \textit{infinite} real state space. The problem that we consider here is to find a switching rule
that selects a mode ensuring that the system will still be
in $V$ at the next sampling time, and so on indefinitely.

Note that, here, we do not impose that the systems always lies
within $V$ between two sampling times, only at sampling times:
if the system goes out of $V$ between two sampling times,
then, due to continuity reasons and because of the ``small'' size of $\tau$,
it will still stay within the close neighborhood of $V$, and we assume
that such a small deviation is acceptable for the system.
This makes the problem simpler than the one considered,
e.g. in \cite{abdmp00}, where the system is forced to always stay
within $V$. 

Note also that the problem here is simpler than the one considered
in \cite{sun-ge-lee02}, because, here, only the switching rule
has to be determined, since the control input is fixed.
(In \cite{sun-ge-lee02}, the dynamics is of the form $\dot{x}(t)=A{x}(t)+Bu(t)$,
where $u(t)$ is not constant, but an input to be synthesized.)

Finally, our problem is much simplified by the fact that,
as in \cite{Girard}, the switching instants can only occur at times of
the form  $i\tau$ with $i\in \mathbb{N}$.

As noted in \cite{abdmp00}, there are two approaches for solving
this kind of problem: 

- the {\em indirect} approach reduces first the system, 
via abstraction, into a discrete event system (typically, a finite-state automaton); this is done in, e.g., \cite{Girard}.
One can thus identify cycles in the graph of the abstract system,
thus inferring possible patterns of modes that enforces the
system to stay forever within $V$.

- the {\em direct} approach works directly on
the continuous state space; this is done, e.g., in \cite{abdmp00}.
One can thus infer a {\em controllable} subspace $V'$ of $V$,
within which the existence of a switching rule allowing
to stay forever within $V'$ is guaranteed (see, e.g., \cite{sun-ge-lee02,krastanov-veliov04}).

Often, in the indirect approach, the switching rule can be computed
{\em off line} (under, e.g., the form of a repeated pattern of modes),
while the switching rule has to be computed {\em on line}
in the direct approach.

Our methods basically rely on previous works, but we specialize them
to the  simplified context 
(linearity, periodic switching instants, absence of control input),
which is motivated by the features of a focused case study:
a DC-DC boost converter built by electronics laboratory SATIE (ENS Cachan) for the automative industry.
Our enhanced methods allow us to treat successfully this real-life example.

\section{Indirect Approach: Approximately Bisimular Methods}
\subsection{Sampled Switched Systems}

In this paper, we consider a subclass of hybrid systems
\cite{Henzinger}, called ``switched systems'' in \cite{Girard}.
\begin{definition}
A {\em switched system} $\Sigma$ is a quadruple $(\mathbb{R}^n,P,\mathcal{P},F)$ where:
\begin{itemize}
\item $\mathbb{R}^n$ is the state space
\item $P=\{1,\dots,m\}$ is a finite set of {\em modes},
\item ${\cal P}$ is a subset of ${\cal S}(\mathbb{R}_{\geq 0},P)$
which denotes the set of piecewise constant functions
from $\mathbb{R}_{\geq 0}$ to $P$, continuous from the right and with a finite number of discontinuities on every bounded interval of
$\mathbb{R}_{>0}$
\item $F=\{f_p\ |\ p\in P\}$ is a collection of functions indexed by $P$.
\end{itemize}
\end{definition}
For all $p\in P$, we denote by $\Sigma_p$ the continuous subsystem 
of $\Sigma$ defined by the differential equation:
$$\dot{{\bf x}}(t)=f_{p}({\bf x}(t)).$$
A {\em switching signal} of $\Sigma$ is a function
${\bf p} \in {\cal P}$, the discontinuities of ${\bf p}$ are called {\em switching times}.
A piecewise ${\cal C}^1$ function ${\bf x}: \mathbb{R}_{>0}\rightarrow \mathbb{R}^n$ is said to be a {\em trajectory} of $\Sigma$ if it is continuous and there exists a switching signal ${\bf p} \in{\cal P}$ such that, at each $t\in \mathbb{R}_{>0}$, ${\bf x}$ is continuously differentiable and satisfies:
$$\dot{{\bf x}}(t)=f_{{\bf p}(t)}({\bf x}(t)).$$ 
We will use ${\bf x}(t,x,{\bf p})$ 
to denote the point reached at time $t\in\mathbb{R}_{>0}$ from the initial condition $x$ under the swiching signal ${\bf x}$.
Let us remark that a trajectory of $\Sigma_p$ is a trajectory of $\Sigma$ associated with the constant signal ${\bf x}(t)=p$, for all
$t\in\mathbb{R}_{>0}$. 

In this paper, we focus on the case of {\em linear} switched systems:
for all $p\in P$, the function $f_p$ is defined by $f_p(x)=A_px+b_p$ where $A_p$ is a $(n\times n)$-matrix of constant elements $(a_{i,j})_p$
and $b_p$ is a $n$-vector of constant elements~$(b_k)_p$.

In the following, as in \cite{Girard}, we will work
with trajectories of duration $\tau$ for some chosen 
$\tau\in\mathbb{R}_{\geq 0}$,
called ``time sampling parameter''. This can be seen as a sampling process.
Particularly, we suppose that switching instants can only occur at
times of the form $i\tau$ with $i \in \mathbb{N}$.
In the following, we will consider transition systems that describe 
trajectories of duration $\tau$, for some given {\em time sampling parameter} $\tau\in\mathbb{R}_{\geq 0}$.
\begin{definition}
Let $\Sigma=(\mathbb{R}^n,P,{\cal P},F)$ be a switched system and $\tau\in\mathbb{R}_{\geq 0}$ a time sampling parameter.
The {\em $\tau$-sampled transition system 
associated to $\Sigma$}, denoted by $T_{\tau}(\Sigma)$ ,
is the transition system $(Q,\rightarrow^p_{\tau})$ defined by:
\begin{itemize}
\item the set of states is $Q=\mathbb{R}^n$
\item the transition relation is given by
$$x\rightarrow^p_{\tau} x' \mbox{ iff } {\bold x}(\tau,x,p)=x'$$
\end{itemize}
\end{definition}
Let us define:
$Post_i(X)=\{x' \ |\ x\rightarrow^i_{\tau} x'  \mbox{ for some $x\in X$}\}$, and\\
$Pre_i(X)=\{x' \ |\ x'\rightarrow^i_{\tau} x \mbox{ for some $x\in X$}\}$.
For the sake of brevity, we will use $Post_i(x)$ instead of $Post_i(\{x\})$ and $Pre_i(x)$ instead of $Pre_i(\{x\})$.

\begin{example}\label{ex:boost1}
This example is a boost DC-DC converter with one switching cell
(see Fig.~\ref{fig:Boost1}) that is taken from \cite{Girard} (see also, e.g., \cite{BecPap:2005:IFA_2345,test,Senesky03hybridmodeling}).
The boost converter has two operation modes depending on the position of the switching cell.
The state of the system is $x(t)=[i_l(t), v_c(t)]^T$ where $i_l(t)$ is
the inductor current and $v_c(t)$ the capacitor voltage. The dynamics 
associated with both modes are of the form $\dot{x}(t)=A_px(t)+b_p$ ($p=1,2$) 
with
\[A_1 = \begin{pmatrix} -\frac{r_l}{x_l} & 0 \\ 0 & -\frac{1}{x_c}\frac{1}{r_0+r_c} \end{pmatrix}\ \ \  b_1=\begin{pmatrix} \frac{v_s}{x_l} \\ 0 \end{pmatrix}\]
\[A_2 = \begin{pmatrix} -\frac{1}{x_l}(r_l + \frac{r_0.r_c}{r_0+r_c}) & -\frac{1}{x_l}\frac{r_0}{r_0+r_c} \\ \frac{1}{x_c}\frac{r_0}{r_0+r_c} & -\frac{1}{x_c}\frac{r_0}{r_0+r_c} \end{pmatrix} \ \ \ b_2= \begin{pmatrix} \frac{v_s}{x_l} \\ 0 \end{pmatrix}\]
It is clear that the boost converter is an example of a switched system.
We will use the numerical values of \cite{Girard}:
$x_c = 70$, $x_l = 3$, $r_c = 0.005$, $r_l = 0.05$, $r_0=1$, $v_s=1$. 
The goal of the boost converter is to regulate the output voltage across the load $r_0$. 
This control problem is usually reformulated as a current reference scheme. 
Then, the goal is to keep the inductor current $i_l(t)$ around a reference value $i_l^{ref}$. This can be done, for instance, by synthesizing
a controller that keeps the state of the switched system in an invariant set centered around the reference value.
An example of switching rule
is illustrated on Fig.~\ref{fig:controle_boost1}. This rule is periodic of period $T_d$: the mode is 2 on $(0, \frac{T_d}{4}]$ and 1 on $(\frac{T_d}{4}, T_d]$. 
 A.

\begin{figure}[!ht]
\centering
\includegraphics[scale = 0.25]{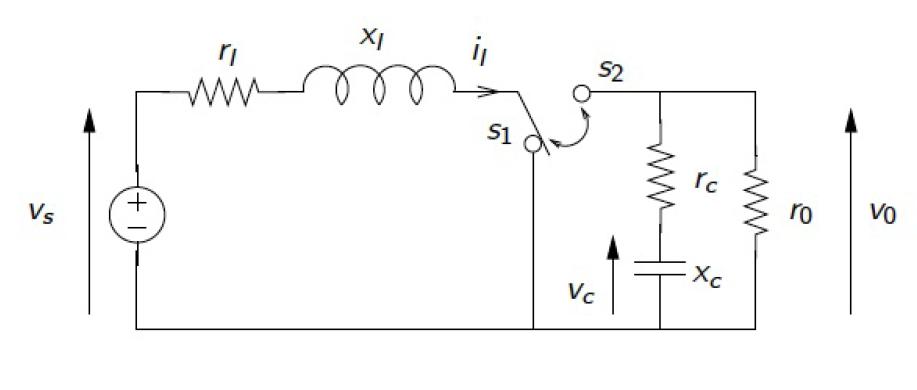}
\caption{Electric scheme of the boost DC-DC converter (1 cell)}
\label{fig:Boost1}
\end{figure}
\begin{figure}[!ht]
\centering
\includegraphics[scale = 0.6]{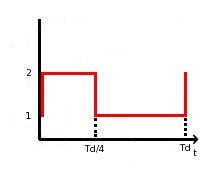}
\caption{A switching rule for the 1-cell boost DC-DC converter on one period of length~$T_d$ (here, $\tau=\frac{T_d}{4}$, and the pattern of the switching rule is $(2.1.1.1)$)}
\label{fig:controle_boost1}
\end{figure}
\end{example}
\subsection{Approximate bisimulation}
In \cite{Girard}, the authors propose  a method for abstracting
a switched system under the form of a discrete symbolic model,
that is equivalent to the original one,
under certain Lyapunov-based stability conditions.
They use an Euclidian metric $\|.\|$, and define
the approximation of the set of states $\mathbb{R}^n$ as follows:
$$[\mathbb{R}^n]_{\eta} = \{x \in \mathbb{R}^n \ | \ x_i = k_i\frac{2\eta}{\sqrt{n}}, \ k_i \in \mathbb{Z}, i=1,\dots,n\},$$ 
where $\eta\in \mathbb{R}_{\geq 0}$ is a state space discretization parameter.
The transition relation $T_{\tau}(\Sigma)$ is approximated as follows:
Let $q \in [\mathbb{R}^n]_{\eta}$ and $q_e = {\bf x}(\tau,q,p)$ such that $q \rightarrow_{T_{\tau}(\Sigma_p)} q_e$ in the real system, let $q' \in [\mathbb{R}^n]_{\eta}$ with  $||q_e - q'|| < \eta$. Then we have $q \rightarrow_{T_{\tau,\eta}(\Sigma_{p})} q'$ 
for the approximated transition relation.
The {\em approximate} transition system $T_{\tau,\eta}(\Sigma)$ 
is defined as follows:
\begin{definition}
The system $T_{\tau,\eta}(\Sigma)$ is the transition system
$(Q,\rightarrow^p_{\tau,\eta})$ defined by:
\begin{itemize}
\item the set of states is $Q=[\mathbb{R}^n]_{\eta}$
\item the transition relation is given by
$$q\rightarrow^p_{\tau,\eta} q' \mbox{ iff } \|{\bold x}(\tau,q,p)-q'\|\le \eta$$
where $\| . \|$ is any metric on $\mathbb{R}^n$. 
\end{itemize}
\end{definition}


The notion of ``approximate bisimilarity'' between
systems
$T_{\tau}(\Sigma)$ and $T_{\tau,\eta}(\Sigma)$ is defined as follows:
\begin{definition}
Systems $T_{\tau}(\Sigma)$ and $T_{\tau,\eta}(\Sigma)$ are {\em $\varepsilon$-bisimilar} if:
\begin{enumerate}
\item $(\| x-q\|\leq\varepsilon \wedge q\rightarrow^p_{\tau,\eta} q') \ \ \Rightarrow\ \ \| x'-q'\|\leq\varepsilon$\\
for some $x'={\bold x}(\tau,x,p)$ (i.e. for some ${\bf x}':$ $x\rightarrow^p_{\tau} x'$), and
\item $(\| x-q\|\leq\varepsilon \wedge x\rightarrow^p_{\tau} x') \ \ \Rightarrow \ \ \| x'-q'\|\leq\varepsilon$\\
for some $q'\in [\mathbb{R}^n]_{\eta} : \|{\bold x}(\tau,q,p)-q'\|\le \eta$ (i.e. for some $q':$ $q\rightarrow^p_{\tau,\eta} q'$).
\end{enumerate}

\end{definition}

The following theorem is given in \cite{Girard}.

\begin{theorem}
Consider a switched system $\Sigma = (\mathbb{R}^{n}, P,{\cal P}, F)$ with
${\cal P} = {\cal S}(\mathbb{R}^{+}, P)$, a desired precision $\varepsilon$ and a time sampling value $\tau$.
Under certain Lyapunov-based stabilization conditions, there exists a space sampling value $\eta$ such that the transition systems
$T_{\tau}(\Sigma)$ and $T_{\tau,\eta}(\Sigma)$ are approximately bisimilar with precision~$\varepsilon$.
\end{theorem}
One can guarantee an arbitrary precision $\varepsilon$ by choosing
an appropriate $\eta$: there exists an explicit algebric relation between $\varepsilon$ and $\eta$. Furthermore, under certain conditions (stability of $T_{\tau}(\Sigma)$), the symbolic model $T_{\tau,\eta}(\Sigma)$ has a finite number of states.
One can then use standard techniques of model checking
in order to synthesize a safe switching rule on $T_{\tau,\eta}(\Sigma)$ (e.g., 
letting the system always in the safe area), see e.g. \cite{AVW03,RW89}.
The switching rule on $T_{\tau,\eta}(\Sigma)$ can also be used to enforce
the real system $T_{\tau}(\Sigma)$ to behave correctly.

\subsection{Simplification for the Case of Linear Dynamics}
By focusing on {\em linear} dynamics,
we are allowed to simplify the more general method of \cite{Girard} as follows:
\begin{enumerate}
 \item  We are using the infinity norm in order to remove the overlapping of two adjacents bowls of radius $\eta$ (reducing it to a set with a norm 0). This is done to prevent non-determinism. Therefore, $[\mathbb{R}^n]_{\eta}$ has to be changed according to the use of this norm. From now on, $[\mathbb{R}^n]_{\eta} = \{x \in \mathbb{R}^n \ | \ x_i = 2k\eta \ k \in \mathbb{Z}\}$
 \item The computation of Lyapunov functions
 are not necessary in our particular case but can be done by simply computing the infinite sum of a geometric serie  to ensure the $\varepsilon$-bisimulation.
Stability criterion relies simply on the eigenvalues of matrices $A_i$ having negative
real part.
The proof of $\varepsilon$-bisimilarity is based on the fact that $\beta_{\tau}\varepsilon+\eta\leq\varepsilon$ (which is true for some $\eta$ when $\beta_{\tau}<1$) (See \cite{rr-lsv-11-12} for more details).
  \item Due to the presence of the exponential of a matrix, the computation of the image of all the points could be very costly. By using the linearity of the system, we can compute the same results for a fraction of the initial cost. This is explained in \cite{rr-lsv-11-12}.
\end{enumerate}
\begin{example}
Our simplified method is applied on the boost converter of Example~\ref{ex:boost1}:
$\tau = 0.5$, $V$ corresponding to $i_l \in [3,3.4]$ and $v_c \in [1.5,1.8]$, $\varepsilon = 3.0$, $\eta = 1/40$.\footnote{The values used are not the same as the ones used by the authors of \cite{Girard} due to a rescaling done in  \cite{Girard}}. See Fig.~\ref{fig:graph_1_40_part} for one of the connected component of the full graph. Each cycle in the graph corresponds to a periodic control of the converter which ensures that the electric variables lie inside the predefined $V$ up to $\varepsilon$. For example, we consider the cycle going through the vertices: 
$159,243,173,257,187,271,201,285,215,299,\\229,159$. This corresponds to the periodic mode control of pattern $12121212122$. The result of a simulation under this periodic switching rule is given in Fig. \ref{depliement} for a starting point $x_0 = (3.0,1.79)$. The box $V$ is delimited by the red lines. 
One can see that the system largely exceeds the limits of $V$ (but stays inside the $\varepsilon$-approximation).

\begin{figure}[!ht]
 \centering
 \includegraphics[scale = 0.3]{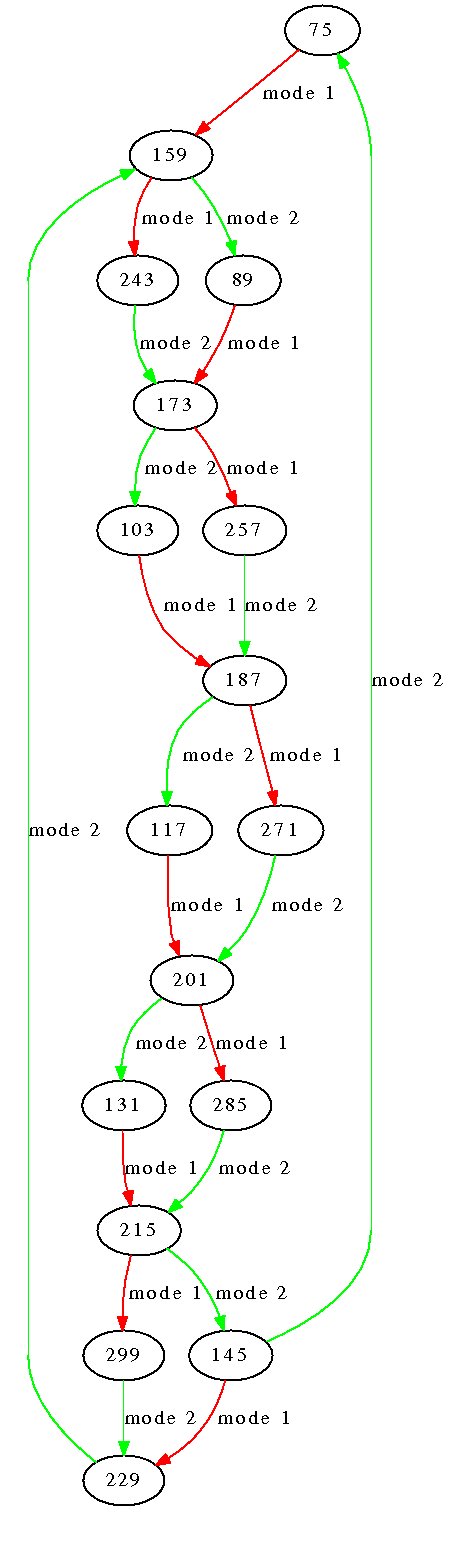}
 \caption{Synthesized finite-state automaton for 1-cell converter with $\eta = \frac{1}{40}$ and $V$ corresponding to $i_l \in [3,3.4]$ and $v_c \in [1.5,1.8]$}
 \label{fig:graph_1_40_part}
\end{figure}


\begin{figure}[h!]
  \centering
  \label{fig:exemple1}
  \includegraphics[scale=0.4]{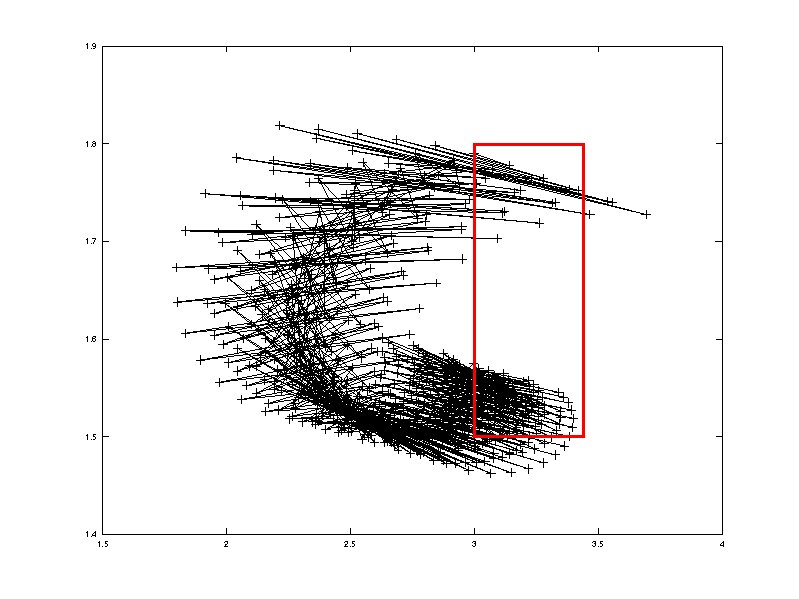}
  \caption{Projected simulation with $i_l$ on the abscissa, and $v_c$ on the ordinates of 1-cell converter for switching rule $(12121212122)^*$ starting point $x_0 = (3.0,1.79)$ ; The box $V = [3,3.4] \times [1.5,1.8]$ is drawn in red. ($\varepsilon =2.6$)}
  \label{depliement}
\end{figure}

%
\end{example}

\section{Direct Approach: Inference of Controllable Subspace}
\label{section_new_method}
The {\em direct} approach works directly on
the continuous state space; this is done, e.g., in \cite{abdmp00}.
One can thus infer a {\em controllable} subspace $V'$ of $V$,
within which the existence of a switching rule allowing
to stay forever within $V'$ is guaranteed (see, e.g., \cite{sun-ge-lee02,krastanov-veliov04}). We present here a simplified direct 
method that exploits the simple features of  our framework:
linearity, absence of perturbation $u$, periodicity of the switching instants.

Consider a box $V\subset \mathbb{R}^n$ and a time sampling 
value $\tau$.
The following Algorithm \ref{algo} computes a set of controllable polyhedra.\
Intuitively, after the $k^{th}$ iteration of the loop, $Control_i$ is a set of states satisfying the following property:
there exists a sequence of modes $\sigma$ of length $k$ starting with mode $i$ such that $\sigma$ applied to any state of $Control_i$ prevents the system to go out of $V$ at any sampling time; alternatively, after the $k^{th}$ iteration, $Uncontrol$ is a set of states for which, 
for all sequence $\sigma$, there exists a prefix $\sigma'$ which makes the system go outside $V$.
Note that the termination of the procedure is not guaranteed due to the fact that there are infinitely many polyhedral sets.\\
 \begin{algorithm}[H]
\SetKwInOut{Input}{input}\SetKwInOut{Output}{output}

\Input{A switched system $\Sigma$ with $m$ modes}
\Input{A time sampling value $\tau$}
\Input{A box $V \subset \mathbb{R}^n$}

\Output{A controllable subspace $V'=\{Control_i\}_{i=1..m}$}
\BlankLine
$Uncontrol_{new} := V$\\
\For{$i=1..m$}{
$Control_i := V$
}
\Repeat{$Uncontrol_{new} = Uncontrol$}{
	$Uncontrol := Uncontrol_{new}$\\
	\For{$i=1..m$}{$Control_i := Control_i \setminus Pre_i(Uncontrol)$}
	$Uncontrol_{new} := V \setminus \bigcup_{i=1}^{m} Control_i$}
\Return{$V'=\bigcup_{i=1}^{m} Control_i$}

%
%

\caption{Synthesis  of controllable subspace}
\label{algo}
\end{algorithm}
The correctness of Algorithm \ref{algo} relies on the following fact:
\begin{theorem}\label{th:2}
 If Algorithm \ref{algo} terminates with output $V'$, then $Post_i(V') \subseteq V'$ for all $1 \leq i \leq m$.
\end{theorem}
In order to prove Theorem \ref{th:2}, we need the two following propositions.

\begin{proposition}
\label{injectivite}
  Let $x,x' \in \mathbb{R}^n$. Then the three following items are equivalent:
\begin{itemize}
 \item $x = x'$
 \item $Post_i(x) = Post_i(x')$ for all $1 \leq i \leq m$
 \item $Pre_i(x) \neq Pre_i(x')$ for all $1 \leq i \leq m$
\end{itemize}
\end{proposition}
\flushleft{\textbf{Proof}}\\
We only consider Linear Differential Equations, for any $x \in \mathbb{R}^n$ $Card(Post_i(x)) = Card(Pre_i(x)) = 1$. The proof is immediate from Cauchy-Lipschitz Theorem. $\hspace*{\fill} \Box$

\begin{proposition}
\label{ensembliste}
Let $A,B \subseteq \mathbb{R}^n$. We have:
\begin{enumerate}
 \item $Post_i(A \setminus B) = Post_i(A) \setminus Post_i(B)$  and similary for $Pre_i$ for all $1 \leq i leq m$.
 \item $Post_i(Pre_i(A)) = Pre_i(Post_i(A)) = A$.
\end{enumerate}

\end{proposition}
\flushleft{\textbf{Proof}}\\
Let $A,B \subseteq \mathbb{R}^n$ and $i \in \{1..m\}$. We have:
\begin{enumerate}
 \item Let $y \in Post_i(A \setminus B)$. Then $x \rightarrow^{i}_{\tau} y$ for some $x \in A \setminus B$. Therefore $y \in Post_i(A)$ since $x \in A$. Let suppose that $y \in Post_i(B)$ then $x' \rightarrow^{i}_{\tau} y$ for some $x' \in B$. By Proposition \ref{injectivite}, $x=x'$ which raises a contradiction. Therefore $y \not \in Post_i(B)$ and $y \in Post_i(A) \setminus Post_i(B)$.\\
Let $y \in Post_i(A) \setminus Post_i(B)$. Therefore, $y \in Post_i(A)$ and $x \rightarrow_{\tau} y$ for some $x \in A$. However, $y \not \in Post_i(B)$, therefore $\forall x' \in B, x' \not \rightarrow_{\tau} y$. It follows that $x \in A \setminus B$. Therefore $y \in Post_i(A \setminus B) \hspace*{\fill} \Box $
 \item We know that $Card(Post_i(Pre_i(x))) =1$ and it is immediate that $x \in Post_i(Pre_i(x))$ therefore $\{x\} = Post_i(Pre_i(x))$. Therefore $Post_i(Pre_i(A)) = A$ and by the same type of proof $Pre_i(Post_i(A))=A. \hspace*{\fill} \Box $
\end{enumerate}

We can now prove Theorem \ref{th:2}
\flushleft{\textbf{Proof of Theorem \ref{th:2}}}\\
We will denote by $Control^k_i$ the value of $Control_i$ at the $k^{th}$ iteration of the algorithm. Suppose that the Algorithm terminates at iteration $n$. Therefore, at iteration $n$, $Uncontrol_{new} = Uncontrol$. Let $x \in Control^n_i$ for a given $i \in \{1..m\}$.Suppose that $Post_i(x) \not \in V'$ then $Post_i(x) \in Uncontrol_{new}$. Therefore $Post_i(x) \in Uncontrol$. By Proposition \ref{ensembliste}, $x \in Pre_i(Uncontrol)$, but we have that $Control^n_i = Control^{n-1}_i \setminus Pre_i(Uncontrol)$. Therefore $x$ does not belong to $Control^n_i$ which raises a contradiction. Therefore $Post_i(x) \in V'$ and $Post_i(V') \subseteq V' \hspace*{\fill} \Box$\\

In other words, the output set $V'$ of controllable polyhedra is invariant. 
Let us point out that the system may temporarily  go out
of $V$ between two sampling instants.

It immediately follows from Theorem \ref{th:2} that, at any sampling
time $i\tau$, any point in
$V'\subset V$, is controllable:
there exists a mode~$j$ that ensures that the point at next sampling time $(i+1)\tau$ is still in $V'$.

Note that, unlike the indirect method, the appropriate mode cannot be 
precomputed, but has to be found {\em on line}.
On the other hand, the system lies {\em exactly} within $V'\subset V$ at 
each sampling
time (instead of lying within the $\varepsilon$-closeness of $V$ using the indirect method).

%
%
%
%
%


Algorithm \ref{algo} involves the computation of the $Pre$-image of 
(union of) convex polyhedra.
We have (see \cite{rr-lsv-11-12} for a proof):
\begin{lemma}
\label{lemme_barycentre}
Let $\tau \in \mathbb{R}_{>0}$ and $S$ a convex set of $\mathbb{R}^{n}$ and $i$ a mode of $\Sigma$. Then\\
 $Pre_i(S)$ is a convex set of $\mathbb{R}^n$.
\end{lemma}
From this Lemma, we can compute the $Pre_i$-image of any convex polyhedron by simply computing the $Pre_i$-image of its vertices. 

Unfortunately, Algorithm \ref{algo} also involves the computation of union, complementation,
and test of equality of polyhedra, that are 
operations known to be very expensive.
To overcome this problem, one can approximate all the manipulated objects
using the notion of {\em griddy polyhedra} (see \cite{abdmp00,bmp99}),
i.e., sets that can be written as unions of closed unit hypercubes
with integer vertices. 
The price to be paid is an underapproximation of the controllability subspace,
but this kind of compromise seems unavoidable,
as pointed out in \cite{abdmp00}.
\begin{example}
To illustrate this approach, we are computing a control for the boost DC-DC converter with one cell, see Example \ref{ex:boost1} for a description of the system. The resulting control presented in Fig. \ref{geom_boost1} has been obtained for the following parameters:
$V$ corresponding to $i_l \in [3.0, 3.4]$ and $v_c \in [1.5,1.8]$, $\tau = 0.5$ and $x_0 = (3.01, 1.79)$.
The Fig. \ref{geom_boost1} can be decomposed into 4 parts:
\begin{itemize}
 \item Two big vertical polyhedra. The left one represents the zone controllable with mode 1, the right one with mode 2.
 \item Two small horizontal polyhedra (upper right and lower left) 
are the uncontrollable zones of $V$.\footnote{They are
delimited by vertices: $(3,1.5);(3.1092,1.5);$
$(3.1110,1.5107);(3.0000,1.5107)$ and $(3.2611,1.7897);(3.4,1.788);$
$(3.4 ,1.8);(3.2611,1.8)$.}
\end{itemize}
A trajectory starting from point $x_0=(3.01,1.79)$
belonging to the controllable subset, and using an on-line computation
of the switching rule,
has been depicted on Fig. \ref{geom_boost1}. and  \ref{geom_boost1bis} for a simulation.
One can see that the trajectory {\em always} stays within $V$
(not only at the sampling instants).
\begin{figure}[!ht]
 \centering
 \includegraphics[scale = 0.60]{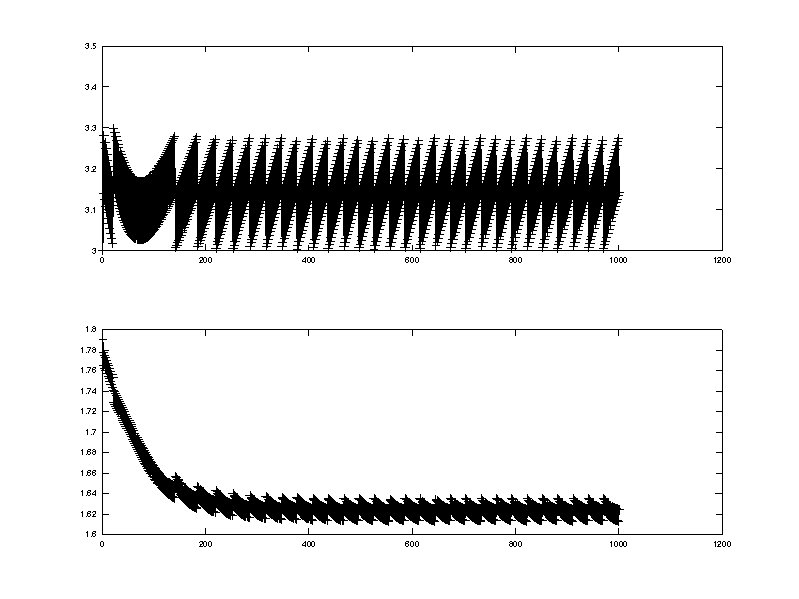}
 \caption{Simulation of the 1-cell boost converter
starting from point $x_0=(3.01, 1.79)$ that belongs
to the controllable subspace of $[3.0,3.4]\times[1.5,1.8]$. Above: simulation of $v_c$ during time; below: simulation of $i_l$}
 \label{geom_boost1bis}
\end{figure}
 \begin{figure}[!ht]
  \centering
  \includegraphics[scale = 0.6]{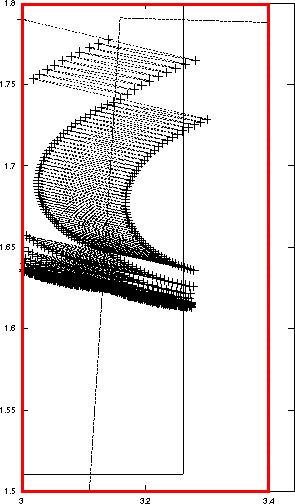}
  \caption{Controllable subspaces of the boost DC-DC converter (1 cell) from $x_0 = (3.01,1.79)$ within $V= [3.0,3.4] \times [1.5,1.8]$}
  \label{geom_boost1}
 \end{figure}


%
\end{example}

\section{Application to a 3-cells DC-DC Boost Converter}

\begin{figure}[!ht]
\centering
\includegraphics[scale = 0.1]{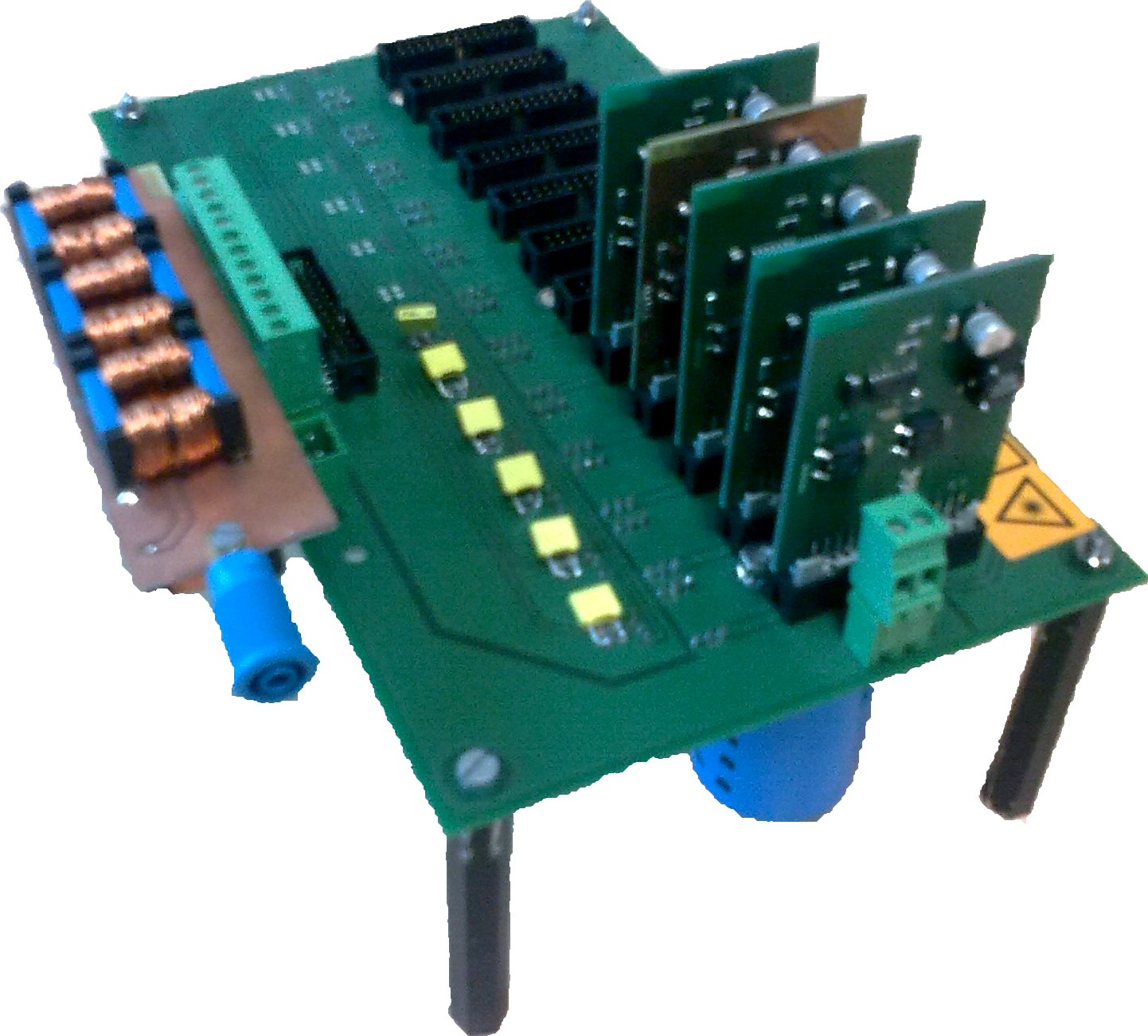}
\caption{3-cells converter built by laboratory SATIE}
\label{fig:boost3bis}
\end{figure}

Our method is scalable to bigger systems as we illustrate with the boost DC-DC converter with 3 cells. This is a real-life system built by the electronics laboratory SATIE (ENS Cachan) for the automative industry. See Fig. \ref{fig:boost3bis} for a picture of the system.
\subsection{Model}
\begin{figure}[!ht]
\centering
\includegraphics[scale = 0.25]{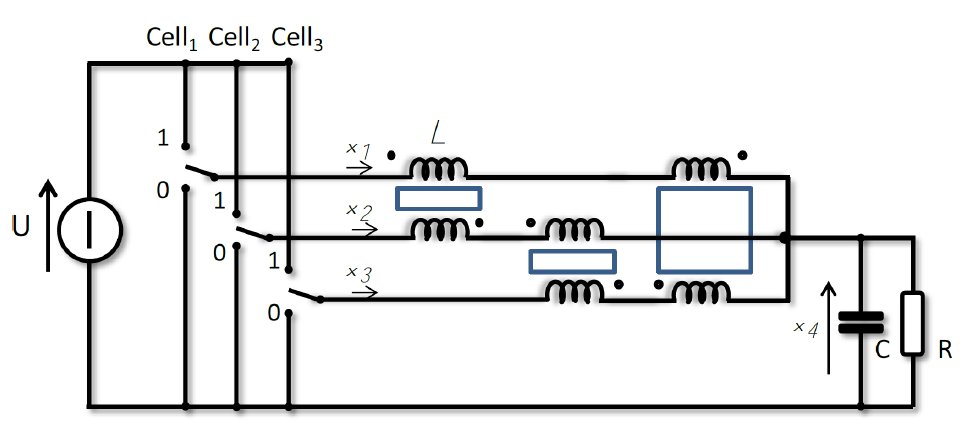}
\caption{Electrical scheme of the DC-DC converter with 3 cells}
\label{fig:boost3}
\end{figure}

\begin{figure}[!ht]
\centering
\includegraphics[scale=0.7]{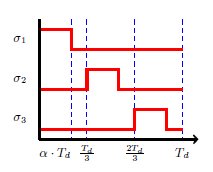}
\caption{
Switching rule for the 3-cells boost DC-DC converter on one period of length~$T_d$ (here, $\alpha=\frac{1}{2}$, $\tau=\frac{T_d}{6}$, 
$\sigma_1=(1.0^5)$, $\sigma_2=(0^2.1.0^3)$, $\sigma_3=(0^4.1.0)$,
and the pattern of the corresponding switching rule is $(2.1.3.1.5.1)$)}
\label{fig:controle_boost3}
\end{figure}

The boost DC-DC converter with 3 cells relies on the same principle as the one with one cell. 
An advantage of this model is its robustness: even if one switching cell
is damaged, the system is still controllable with the restricted set
of modes that remain available. This system is naturally more complex:
There are 4 continuous variables of interest (instead of two),
and $2^3=8$ modes (instead of two).
The electrical scheme is presented in Fig. \ref{fig:boost3}. An example of pattern of periodic switching rule is presented in Fig. \ref{fig:controle_boost3}.

The system satisfies the following equations:\\
\[U \begin{pmatrix} \sigma_1 \\ \sigma_2 \\ \sigma_3 \\ 0 \end{pmatrix} + \begin{pmatrix} -2r & 0 & 0 & -1 \\ 0 & -2r & 0 & -1 \\ 0 & 0 & -2r & -1 \\ 1 & 1 & 1 & -1/R \end{pmatrix}. \begin{pmatrix} x_1 \\ x_2 \\ x_3 \\ x_4 \end{pmatrix} = \begin{pmatrix} 2L & -M & -M & 0 \\ -M & 2L & -M & 0 \\ -M & -M & 2L & 0 \\ 0 & 0 & 0 & C \end{pmatrix}.\frac{d}{dt} \begin{pmatrix} x_1 \\ x_2 \\ x_3 \\ x_4 \end{pmatrix}\]
That can be rewritten to fit our framework as:\\
\[\dot{ x} = M_{LC}^{-1}M_Sx + B_{\sigma}\]
with\\ $M_{LC} = \begin{pmatrix} 2L & -M & -M & 0 \\ -M & 2L & -M & 0 \\ -M & -M & 2L & 0 \\ 0 & 0 & 0 & C \end{pmatrix}$, $M_S = \begin{pmatrix} -2r & 0 & 0 & -1 \\ 0 & -2r & 0 & -1 \\ 0 & 0 & -2r & -1 \\ 1 & 1 & 1 & -1/R \end{pmatrix}$, $B_{\sigma} = U M_{LC}^{-1}.\begin{pmatrix} \sigma_1 \\ \sigma_2 \\ \sigma_3 \\ 0 \end{pmatrix}$\\
where $U$ is the input voltage (here $U = 100$).
\subsection{Indirect Method}
\begin{figure}[!ht]
 \centering
 \includegraphics[scale = 0.25]{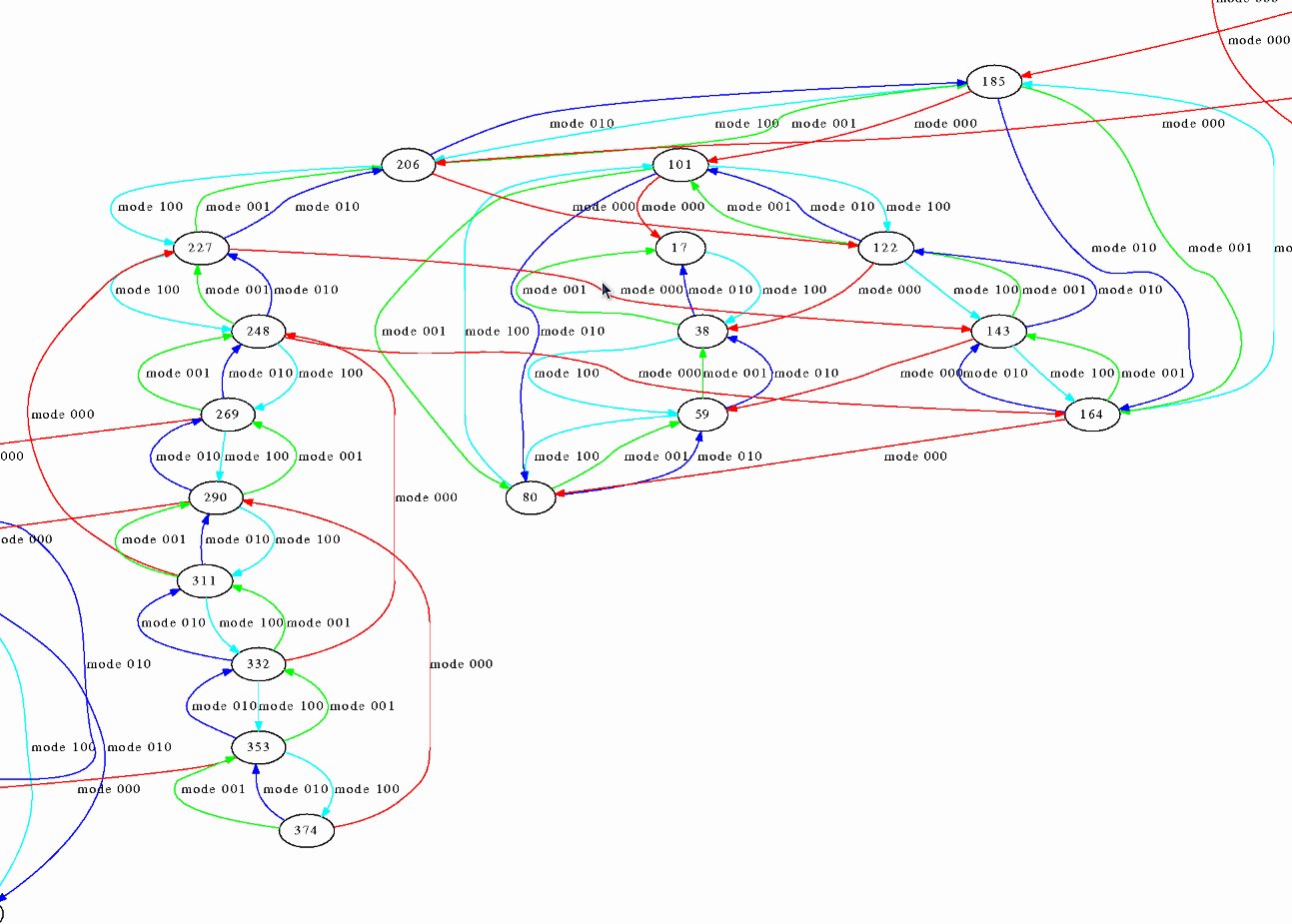}
 \caption{A part of the abstract graph synthesized for the 3-cell converter
of the 3-cells boost converter using the indirect method}
 \label{fig:pattern}
\end{figure}

Here are the parameters that we used: $\eta = 1/5$ which corresponds to $\varepsilon = 21.6$, $\alpha = \frac{1}{2}$, $T_d= 1/10000$, $\tau = 1/60000$. $V$ is defined by: $[5.3,5.9]\times [5.3,5.9]\times [5.3,5.9]\times [15.5,16.5]$.\\
Over a period $T_d=6\tau$,
the switching rule 
(see Fig. \ref{fig:controle_boost3}) 
corresponds to: $100000$ for $\sigma_1$, $001000$ for $\sigma_2$ and $000010$ for $\sigma_3$, which can be represented by the global
pattern $(100,000,010,000,001,000)$
The abstract system for box $V$ 
corresponds to a big graph composed  of many repeated pattern:
a small part of the full graph is given in Fig. \ref{fig:pattern}.
A typical cycle can be seen through states $290,311,332,353,332,311,290$ this corresponds to the pattern modes $444121$.
The construction of the full graph (including some optimizations, like
the deletion of vertices
from which every rule leads to a deadend) took less than 2 minutes. 

\begin{figure}[!ht]
 \centering
 \includegraphics[scale = 0.6]{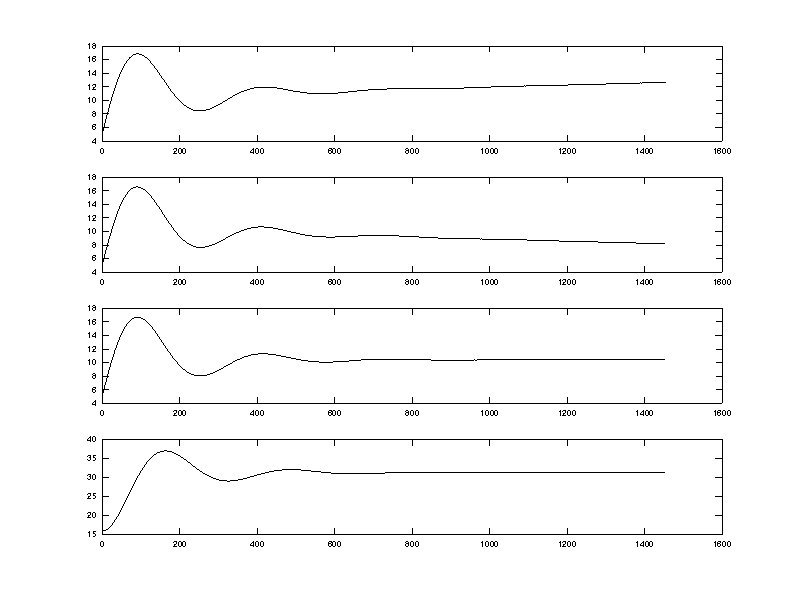}
 \caption{Simulation of the 3-cells Converter under Switching Rule  $(333010)^*$}
 \label{fig:control444121}
\end{figure}

From this graph, we extracted several cycles
that correspond to different switching rules.
We  have simulated the system starting from point
$x_0= (5.4,5.4,5.4,16)$ for such various rules.
The result of one simulation 
for one of them (viz., $(444121)^*$)
is given in Fig. \ref{fig:control444121}.
We can see that, under all these controls, the system goes out of
the initial $V$. However we can check that the system 
stays within the $\varepsilon$-overapproximation of $V$ with $\varepsilon = 21.6$. 
Let us point out incidentally that such an $\varepsilon$
is much too gross to guarantee a realistic precision.
Rather than presenting the results of the indirect approach
with a better precision (using a finer $\eta$-grid),
we present henceforth the results obtained with the direct approach.

\subsection{Direct Method}

\begin{figure}[!ht]
 \centering
 \includegraphics[scale = 0.6]{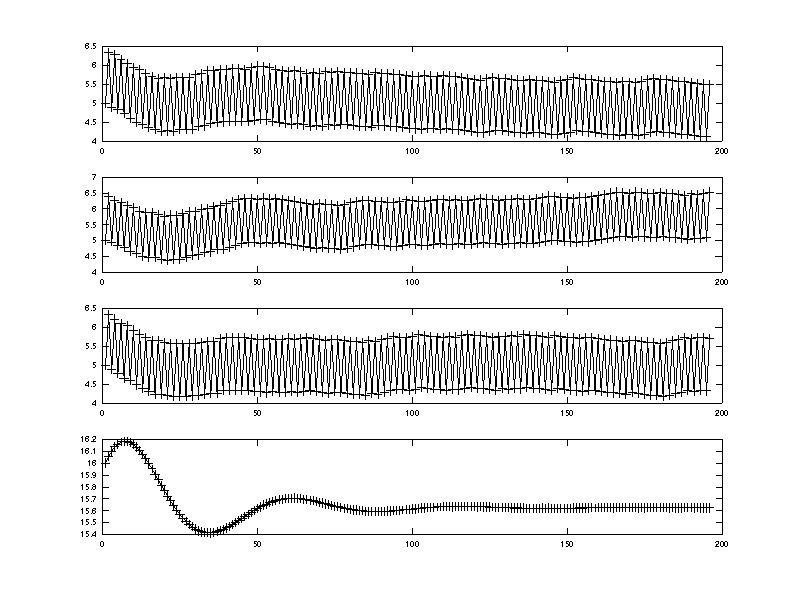}
 \caption{Simulation of 3-cells converter starting from $x_0 = (5,5,5,16)$ in $V = [4,7] \times [4,7] \times [4,7] \times [15,17]$
(from top to bottom: $x_1,x_2,x_3,x_4$ in function of time) }
 \label{fig:geom_boost3}
\end{figure}

\begin{figure}[!ht]
  \centering
  \includegraphics[scale = 0.6]{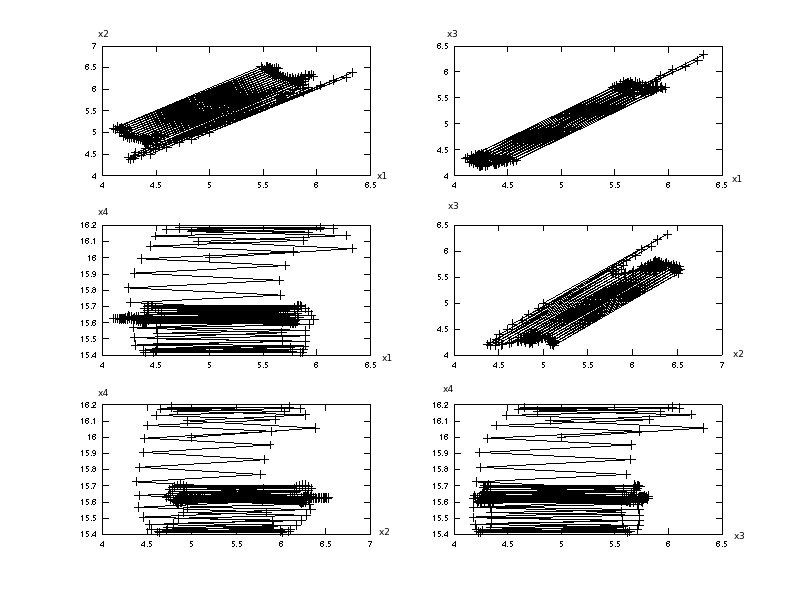}
  \caption{Various projections on plans $(x_i,x_j)$ (for $i,j\in\{1,...,4\}$)
 of simulations
 of the 3-cells boost converter starting from $x_0 = (5,5,5,16)$ in $V = [4,7] \times [4,7] \times [4,7] \times [15,17]$}
  \label{fig:geom_boost3_phase}
 \end{figure}

For $V = [4,7] \times [4,7] \times [4,7] \times [15,17]$, $\tau = 1/60000$, we can extract a controllable subspace $V' \subset V$. A simulation of the system starting from $x_0=  (5,5,5,16) \in V'$ is presented in 
Fig.~\ref{fig:geom_boost3} (see also Fig.~\ref{fig:geom_boost3_phase}
for a projected simulation). We can check on the figure
that all the simulation lies within $V$.

\subsection{Control on failure}
We have also experimented the methods in a case of failure of
one switching cell of the 3-cells boost converter: we have supposed that
the cell 1 is stuck on position open ($\sigma_1 = 0$),
which means that only 4 of the 8 modes are still available.
The description of our experiments is beyond the scope of this paper.
Let us just point out that we were able to find a switching rule 
for this downgraded
context using the {\em direct} method, but not with the
indirect approach.
\section{Final Remarks}
We have explained how to improve two methods (the direct and indirect ones)
for synthesizing control of a piecewise linear system
by exploiting the special features of a framework met in the case
of a real-case example.
Our experiments show that the advantage of the indirect method
is to allow the user to precompute a periodic control rule
at the price of a certain loss of precision.
On the other hand, the direct method relies on a on-line computation of the
switching rule, but allows us
to satisfy {\em exact} reachability invariance properties.
Furthermore, the direct method seems to be able to treat more easily
limit cases where the system works in a downgraded configuration
due to a failure of one its components.
\section*{Acknowledgment}
We are grateful to Laurent Doyen for his helpful comments on an earlier draft of this paper.
\nocite{*}
\bibliographystyle{eptcs}
\bibliography{generic}

\end{document}